\documentclass[twocolumn,showpacs,preprintnumbers,amsmath,amssymb]{revtex4}
\usepackage{graphicx}
\usepackage{dcolumn}
\usepackage{bm}
\begin{document}
\title {The on-top pair-correlation density 
in the homogeneous electron liquid}
\author {Zhixin Qian}
\affiliation{Department of Physics,
Peking University, Beijing 100871, China}
\date{\today}
\begin{abstract}
The ladder theory, in which the Bethe-Goldstone equation
for the effective potential between two scattering particles
plays a central role,
is well known for its
satisfactory description
of the short-range correlations 
in the homogeneous electron liquid.
By solving exactly the Bethe-Goldstone
equation in the limit of large transfer momentum between
two scattering particles, we obtain accurate results for the on-top
pair-correlation density
$g(0)$, in both three dimensions
and two dimensions.
Furthermore, we prove, in general, the ladder theory
satisfies the
cusp condition for the pair-correlation density $g(r)$ at
zero distance $r=0$.

\end{abstract}
\maketitle

\section{Introduction}

The pair-correlation density $g(r)$ is one of the key concepts in
describing the correlation effects, arising from
Pauli exclusion principle and Coulomb interaction, in
the homogeneous electron liquid (or gas).\cite{ichimaru}
It also plays a significant role
in the constructions of the exchange-correlation
energy density functionals in density-functional theory (DFT),\cite{HK}
since in such constructions the homogeneous electron system is
conventionally taken as a reference system. A great deal
of theoretical progress has recently been made in giving
an accurate evaluation of $g(r)$, or the more specific
spin-resolved pair-correlation densities $g_{\sigma \sigma'}(r)$,
with $g(r)= \frac{1}{2} [ g_{\uparrow \downarrow}(r)
+ g_{\uparrow \uparrow}
(r) ]$.\cite{wang,bachelet,gori-giorgi1,gori-giorgi2,polini,davoudi,nagy}
In particular,
$g(0)$, the on-top pair-correlation density, 
which arises totally from $g_{\uparrow \downarrow}(0)$
since $g_{\uparrow \uparrow}(0) =0$,
has been well known to play a special role in
DFT.\cite{burke} The important implication of $g(0)$ was also
realized in many-body theory long ago because the random phase
approximation (RPA),\cite{mahan} due to its lack
of accurate description of the short-range electron correlations, yields
erroneous negative values for $g(0)$ when the
electron densities are not sufficiently high.\cite{glick-ueda}

It is well known that, in many-body theory, 
the long-range correlations can be rather
successfully taken into account in the RPA, while
the short-range correlations can be properly
described by the ladder theory (LT).\cite{hede,yasuhara1,yasuhara2,brown}
In this paper, we attempt to investigate the short-range
correlations in terms of $g (0)$ in the LT, in both
three dimensions (3D) and two dimensions (2D).
In fact, investigations in this direction date back long ago,
and a great deal of achievement
has been made.
\cite{hede,yasuhara1,yasuhara2,brown,ousaka,nagano,takayanagi,calmels,luis,tanatar}

It is necessary here to give some introduction to the LT.
The effective interaction $V_{eff} ({\bf p}, {\bf p}'; {\bf q})$
in the LT between two scattering electrons with respective momenta
${\bf p}$ and ${\bf p}'$ satisfies
the following Bethe-Goldstone equation:\cite{bethe}
\begin{eqnarray}  \label{BG1}
&& V_{eff} ({\bf p}, {\bf p}'; {\bf q}) = v(q)
+\sum_{\bf k} v({\bf q} - {\bf k})   \nonumber \\
&& \times \frac{(1-n({\bf p}+{\bf k}))(1- n({\bf p}' - {\bf k}))}
{\epsilon_p + \epsilon_{p'} - \epsilon_{{\bf p} + {\bf k}}
- \epsilon_{{\bf p}' - {\bf k}}}
V_{eff} ({\bf p}, {\bf p}'; {\bf k}),
\end{eqnarray}
where $v(q)$ is the Fourier transform of the
Coulomb potential, $n({\bf p}) = \theta(k_F -p)$ is the
momentum distribution in the noninteracting
ground state and $k_F$ is the Fermi momentum, and
$\epsilon_p= \hbar^2 p^2/2m$.

As mentioned above, the RPA gives poor description of the short-range
correlations of the electrons, especially for $g(r)$
as $r \to 0$. In fact, the results for $g_{\uparrow \downarrow}(r)$ in
the RPA violate the following cusp condition:
\cite{yasuhara1,ousaka,kimball,rajagopal,rassolov,sahni}
\begin{eqnarray}  \label{cusp}
\frac{\partial g_{\uparrow \downarrow} (r)}{\partial r}|_{r=0}
= \frac{2}{(d-1)a_B} g_{\uparrow \downarrow}(0) ,
\end{eqnarray}
where $d=3, 2$ is the number of spatial dimensions, and
$a_B$ is the Bohr radius.
It was shown recently 
\cite{pople} that the pair-correlation
density obtained from the first order perturbation
calculation does not satisfy the cusp condition either.
In this paper, 
we prove that $g_{\uparrow \downarrow} (r)$ calculated
from $V_{eff}({\bf p}, {\bf p}'; {\bf q})$ of
Eq. (\ref{BG1}) satisfies
the cusp condition. This indicates the reliablity of the LT
in the calculations of the pair-correlation density at short range.

The short-range structure of the pair-correlation density is
determined by the behavior of the effective potential
$V_{eff} ({\bf p}, {\bf p}'; {\bf q})$ at large
momentum transfer ${\bf q}$. In the limiting case,
one therefore can approximately 
replace the momenta of the scattering electrons
by zero in Eq. (\ref{BG1}),
\begin{eqnarray}  \label{BG2}
V_{eff} ({\bf 0}, {\bf 0}; {\bf q}) =
v(q) - && \sum_{\bf k} v({\bf q} - {\bf k})  \nonumber \\
&& \times \frac{1 - n({\bf k})}{2 \epsilon_k}
V_{eff} ({\bf 0}, {\bf 0}; {\bf k}).   
\end{eqnarray}
A frequently used approach to solving Eq. (\ref{BG2}) in
the literature is making the following approximation in the
Coulomb kernel in the 
momentum summation:\cite{yasuhara2,nagano,luis,yasuhara3,isawa}
\begin{eqnarray}   \label{kernel}
v({\bf q} - {\bf k}) &=&
 v(q),  ~~~~ q > k    \nonumber \\
&=& v(k) ,  ~~~~ k > q .
\end{eqnarray}
With the preceding approximation, an analytical solution for
$V_{eff} ({\bf 0}, {\bf 0}; {\bf q})$ was
obtained which yields the following well-known result
for $g_{\uparrow \downarrow}(0)$ in 3D, \cite{yasuhara2,yasuhara3,isawa}
\begin{eqnarray}   \label{yasuhara}
g_{\uparrow \downarrow}(0) = \biggl [\sqrt{2 \lambda_3}
/I_1(\sqrt{8 \lambda_3}) \biggr ]^2 ,
\end{eqnarray}
where $\lambda_3=2\alpha r_s/\pi$ with $\alpha=(4/9 \pi)^{1/3}$ and
$r_s=(3 /4 \pi n)^{1/3}/a_B$.
A similar result was obtained in 2D,\cite{nagano}
\begin{eqnarray}   \label{nagano}
g_{\uparrow \downarrow}(0)=\biggl 
[ I_0(\sqrt{4 \lambda_2} \biggr ]^{-2},
\end{eqnarray}
where 
$\lambda_2 =r_s/\sqrt{2}$ with $r_s= 1/\sqrt{\pi n} a_B$ in 2D. 
In Eqs. (\ref{yasuhara}) and
(\ref{nagano}), $I_n(x)$ is the $n$th order
modified Bessel function.

In this paper we have managed to 
solve exactly Eq. (\ref{BG2}), i.e., without
making the approximation of Eq. (\ref{kernel}). Our results for
$g_{\uparrow \downarrow}(0)$
are
\begin{eqnarray}   \label{qian3D}
g_{\uparrow \downarrow}(0) =  \biggl [
\frac{45 (45 + 24 \lambda_3 + 4 \lambda_3^2 )}
{2025 +3105 \lambda_3 + 1512 \lambda_3^2
+ 256 \lambda_3^3 } \biggr ]^2 ,
\end{eqnarray}
in 3D, and
\begin{eqnarray}   \label{qian2D}
g_{\uparrow \downarrow}(0) =  \biggl [
\frac{15 ( 64+25 \lambda_2 + 3 \lambda_2^2 )}
{960+1335 \lambda_2
+ 509 \lambda_2^2 + 64 \lambda_2^3} \biggr ]^2 ,
\end{eqnarray}
in 2D. Equations (\ref{qian3D}) and (\ref{qian2D}) are the main
results of this paper.

The paper is organized as follows: 
In Sect. II, we solve Eq. (\ref{BG2}) exactly both in 3D and 2D.
In Sect. III,
we derive analytically 
the expressions of Eqs. (\ref{qian3D}) and (\ref{qian2D})
for $g_{\uparrow \downarrow} (0)$.
We then compare our results with previous ones in the literature
in Sect. IV. Sect. V is devoted to conclusions.
Some technical points on the solutions for
the coefficients of the large momentum expansions of the
effective potentials are given in Appendix A.
In Appendix B, we prove the cusp condition in the LT.

\section{Exact solution to the Bethe-Goldstone integral equation
at large transfer momentum}

In this section, we present our solution 
to Eq. (\ref{BG2}) at large momentum
transfer ${\bf q}$ in the effective potential in both 3D and 2D.
To this end, we denote $V_{eff} ({\bf 0}, {\bf 0}; {\bf q})$
as $V_{eff}(q)$,
and reduce the momenta with unit $k_F$,
and potentials with $v(k_F)$, respectively.
We present our solution for the 3D case in subsection A, and the 2D
case in subsection B, separately.

\subsection{3D}

After carrying out the angular integrations in the summation of 
the momentum ${\bf k}$, Eq. (\ref{BG2}) becomes
\begin{eqnarray} \label{BGsolution}
V_{eff} (q) = \frac{1}{q^2} - \frac{\lambda_3}{2}
\int^\infty_1 dk V_{eff}(k) \frac{1}{qk} \ln |\frac{q+k}
{q-k}| .
\end{eqnarray}
We expand $V_{eff}(q)$ in the powers of $1/q$,
\begin{eqnarray}   \label{Veffsolution}
V_{eff}(q) = \sum_{n=0}^\infty \frac{a_n}{q^{2n+2}} .
\end{eqnarray}
It can be easily confirmed by iteration that 
no odd power terms in the expansion of $V_{eff}(q)$
exist in the solution to Eq. (\ref{BGsolution}). 
The erroneous odd power terms
introduced into
$V_{eff}(q)$ in Refs. \cite{yasuhara2,yasuhara3,isawa} are purely 
due to the approximation made 
in the Coulomb kernel in Eq. (\ref{kernel}).
We substitute Eq.  (\ref{Veffsolution}) 
into Eq. (\ref{BGsolution}), and obtain
\begin{eqnarray}    \label{M2n+1}
\sum_{n=0}^{\infty} \frac{a_n}{q^{2n+2}} =
\frac{1}{q^2} - \frac{\lambda_3}{2q} \sum_{n=0}
^{\infty} a_n M_{2n+3} ,
\end{eqnarray}
where
\begin{eqnarray}  \label{M2n+1M}
M_{2n+3}(q) = \int_1^\infty dk 
\frac{1}{k^{2n+3}} \ln |\frac{q+k}{q-k}|, ~~~
n \ge 0  .
\end{eqnarray}
By carrying through partial integration on the right
hand side of Eq. (\ref{M2n+1M}), one has,
\begin{eqnarray}  \label{M2n+1MM}
M_{2n+3}(q)=\frac{1}{2n+2} \biggl [
\ln\frac{q+1}{q-1} - 2q \Phi_{n+1}(q) \biggr ],
\end{eqnarray}
where
\begin{eqnarray}
\Phi_{n+1}(q)= \int_1^\infty dk
\frac{1}{k^{2n+2}} \frac{1}{k^2- q^2}.
\end{eqnarray}
$\Phi_{n+1}(q)$ defined in the preceding equation can be evaluated to be
\begin{eqnarray}  \label{Phi}
\Phi_{n+1}(q)= && -\sum_{m=0}^n \frac{1}{q^{2m+2}}
\frac{1}{2(n-m)+1}   \nonumber \\
&& + \frac{1}{2q^{2n+3}} \ln \frac{q+1}{q-1}.
\end{eqnarray}
Substituting Eq. (\ref{Phi}) into Eq. (\ref{M2n+1MM}) yields,
\begin{eqnarray}  \label{M2n+3}
 M_{2n+3}(q) && =  \frac{1}{n+1} \biggl [
\sum_{m=0}^{n} \frac{1}{q^{2m+1}} \frac{1}{2(n-m)+1}  \nonumber \\
&& + \frac{1}{2} \biggl ( \frac{1}{q^{2n+2}} -1 \biggr )
\ln \frac{q -1}{q +1} \biggr ], ~
n \ge 0 .
\end{eqnarray}

Finally, substituting Eq. (\ref{M2n+3}) 
into Eq. (\ref{M2n+1}), and comparing the same power orders
of $1/q$, one obtains the following equations for $a_n$:
\begin{eqnarray}   \label{a0equ}
a_0 = 1 - \lambda_3 \sum_{n=0}^\infty \frac{a_n}{2n +1} ,
\end{eqnarray}
and
\begin{eqnarray}  \label{anequ}
a_n = -\frac{\lambda_3}{2n+1} 
\sum_{l=0}^{\infty} \frac{a_l}{2 (l -n ) +1}
, ~ n \ge 1 .
\end{eqnarray}

Equations (\ref{a0equ}) and (\ref{anequ}) for
$a_n$ can be solved 
exactly in principle. In fact, by making the truncation of $a_n = 0$ for
$n \ge 3$, 
a nearly exact solution can be 
obtained as
\begin{eqnarray}  \label{a0}
a_0 = \frac{45( 45 + 24 \lambda_3 + 4 \lambda_3^2)}{D_3} ,
\end{eqnarray}
\begin{eqnarray}   \label{a1}
a_1 = \frac{15 \lambda_3 (45 + 8 \lambda_3)}{D_3} ,
\end{eqnarray}
and
\begin{eqnarray}   \label{a2}
a_2= \frac{45 \lambda_3 (3+ 4 \lambda_3)}{D_3} ,
\end{eqnarray}
where
\begin{eqnarray}
D_3 = 2025 +3105 \lambda_3 +1512 \lambda_3^2 +256 \lambda_3^3 .
\end{eqnarray}

In Appendix A, we show that the preceding solution for $a_0$,
which is directly related to 
$g_{\uparrow \downarrow}(0)$, as shown in the next section,
is very close to the exact numerical solution 
to Eqs. (\ref{a0equ}) and (\ref{anequ}). In fact,
the large  momentum behavior of $V_{eff}(q)$ is dominated by the
leading terms in the large $q$ expansion of $V_{eff}(q)$
in Eq. (\ref{Veffsolution}), and hence a truncation solution
like the preceding one is almost exact.

\subsection{2D}

In 2D, we make use of the following expression,
\begin{eqnarray}
\frac{2 \pi}{|{\bf q} -{\bf k}|} = \int d {\bf r}
e^{i ({\bf q} - {\bf k}) \cdot {\bf r}} \frac{1}{r},
\end{eqnarray}
and rewrite Eq. (\ref{BG2}) as follows:
\begin{eqnarray} 
V_{eff} (q) = \frac{1}{q} - \frac{\lambda_2}{(2\pi)^2}
&& \int d {\bf k} \theta (k-1) \frac{1}{k^2} V_{eff} (k)  \nonumber \\
&& \times \int d {\bf r} 
e^{i ({\bf q} - {\bf k}) \cdot {\bf r}} \frac{1}{r} .
\end{eqnarray}
Carrying out the angular integrations of ${\bf k}$ and
${\bf r}$, we have
\begin{eqnarray} \label{BGsolution2}
V_{eff} (q) = \frac{1}{q} -
\lambda_2 \int_0^\infty dr \int_1^\infty && dk \frac{1}{k}
V_{eff} (k) \nonumber \\
&& \times J_0 (qr) J_0 (kr) ,
\end{eqnarray}
where $J_n(x)$ is the $n$th order Bessel function.
We expand $V_{eff}(q)$ in the powers of $1/q$ as follows:
\begin{eqnarray}   \label{Veffsolution2}
V_{eff}(q) = \sum_{n=0}^\infty \frac{c_n}{q^{2n+1}} .
\end{eqnarray} 
No even power terms
exist in the solution to Eq. (\ref{BGsolution2}).
Again, the erroneous even power terms \cite{nagano}
appear in $V_{eff}(q)$ due to the 
approximation made in Eq. (\ref{kernel}).

We substitute Eq.  (\ref{Veffsolution2})
into Eq. (\ref{BGsolution2}), and obtain
\begin{eqnarray}    \label{N2n+2}
\sum_{n=0}^{\infty} \frac{c_n}{q^{2n+1}} =
\frac{1}{q} - \frac{\lambda_2}{q} \sum_{n=1}
^{\infty} c_{n-1} N_{2n}(q) ,
\end{eqnarray}
where
\begin{eqnarray}  \label{N2nN}
N_{2n}(q)= \int_0^\infty dx J_0(x) 
\int_1^\infty dk \frac{1}{k^{2n}} J_0(kx/q) .
\end{eqnarray}
Carrying out the integration over $k$ in Eq. (\ref{N2nN}),
one obtains,
\begin{eqnarray}   \label{j0}
N_{2n}(q) && =  \sum_{m=1}^{n} (-2)^{m-1}
\frac{(n-1)!}{(n-m)!}    \nonumber \\
&& \times \int_0^\infty dx 
J_0(x/q) J_m(x)/x^m , ~ n \ge 1.
\end{eqnarray}
The integral on the right hand side of Eq. (\ref{j0})
can be expressed in terms of the hypergeometric function
as follows,\cite{grad}
\begin{eqnarray}
\int_0^\infty dx 
J_0(x/q) && J_m(x)/x^m 
= \frac{\Gamma(\frac{1}{2})}{2^m \Gamma(1) 
\Gamma(m+ \frac{1}{2})}   \nonumber \\
\times && F \biggl (\frac{1}{2}, 
-m+\frac{1}{2}; 1; \frac{1}{q^2} \biggr ), ~ n \ge 1,
\end{eqnarray}
where $\Gamma(\alpha)$
is the Gamma function. Therefore,
one has,
\begin{eqnarray}  \label{N2n}
N_{2n}(q)= \sum_{m=1}^{n} && (-2)^{m-1}
\frac{(n-1)!}{(n-m)! (2m -1)!!}   \nonumber \\ 
\times && F \biggl (
\frac{1}{2}, -m + \frac{1}{2}; 1; 
\frac{1}{q^2} \biggr ), ~n \ge 1 .
\end{eqnarray} 
Substituting Eq. (\ref{N2n}) into Eq. (\ref{N2n+2}), 
and comparing the same power orders
of $1/q$, one finally gets
\begin{eqnarray}   \label{c0equ}
c_0 = 1 - \lambda_2 \sum_{n=0}^\infty \frac{c_n}{2n +1} ,
\end{eqnarray}
and
\begin{eqnarray}  \label{cnequ}
c_n = && \lambda_2 (-)^{n-1} \frac{(2n)!}{2^{3n} n!^3}
\sum_{l=0}^\infty c_l l!
\sum_{m=0}^l (-2)^m   \nonumber \\
&& \times \frac{(2m +1)(2m-1) \dots (2m - 2n +3)}
{(l-m)! (2m +1)!!},
\end{eqnarray}
for $n \ge 1$.

Similarly to the 3D case, Eqs. (\ref{c0equ}) and (\ref{cnequ}) 
can be solved exactly in principle. In fact, a
nearly exact solution can be
obtained as follows by the truncation of $c_n = 0$ for
$n \ge 3$:
\begin{eqnarray}  \label{c0}
c_0 = \frac{15(64 + 25 \lambda_2 + 3 \lambda_2^2)}{D_2} ,
\end{eqnarray}
\begin{eqnarray}   \label{c1}
c_1 = \frac{30 \lambda_2 (8 + \lambda_2)}{D_2} ,
\end{eqnarray}
and
\begin{eqnarray}   \label{c2}
c_2= \frac{45 \lambda_2 (1 + \lambda_2 )}{D_2} ,
\end{eqnarray}
where
\begin{eqnarray}
D_2 = 960 + 1335 \lambda_2 + 509 \lambda_2^2 + 64 \lambda_2^3 .
\end{eqnarray}

\section{Results for $g_{\uparrow \downarrow}(r)$ at small $r$}

The spin-antiparallel pair-correlation density in the LT
can be shown to be \cite{yasuhara2,nagano}
\begin{eqnarray}   \label{gud1}
g_{\uparrow \downarrow} (r)
=\frac{4}{n^2} {\sum_{{\bf p}, {\bf p}'}}' | 1 +
\sum_{\bf q} && D({\bf p}, {\bf p}'; {\bf q})   \nonumber  \\
\times && V_{eff}({\bf p}, {\bf p}'; {\bf q})
e^{i {\bf q} \cdot {\bf r} } |^2 ,  
\end{eqnarray}
where the prime on the summations over
${\bf p}$, ${\bf p}'$ means
the restrictions $0 \le p, p' \le k_F$, 
and $D({\bf p}, {\bf p}'; {\bf q})$
is defined as,
\begin{eqnarray}   \label{D}
D({\bf p}, {\bf p}'; {\bf q}) =
\frac{(1-n({\bf p}+{\bf q}))(1- n({\bf p}' - {\bf q}))}
{\epsilon_p + \epsilon_{p'} - \epsilon_{{\bf p} + {\bf q}}
- \epsilon_{{\bf p}' - {\bf q}}} .
\end{eqnarray}

Below we present the results for the 3D and 2D cases
in subsection A and B, respectively. We will reduce
$r$ with unit $1/k_F$.

\subsection{3D}

Using the approximate solution $V_{eff} (q)$ for 
$V_{eff} ({\bf p}, {\bf p}'; {\bf q})$, one obtains \cite{ousaka}
\begin{eqnarray}   \label{gud(r)}
g_{\uparrow \downarrow} (r)
= \biggl [ 1 - \lambda_3 \int_1^\infty  
d q V_{eff} (q) j_0(qr) \biggr ]^2 .
\end{eqnarray}
Trivially,
\begin{eqnarray}
g_{\uparrow \downarrow} (0)
= \biggl [ 1 - \lambda_3 \int_1^\infty  d q V_{eff} (q) \biggr ]^2 .
\end{eqnarray}
With the expression of Eq. (\ref{Veffsolution}),
one has
\begin{eqnarray}
g_{\uparrow \downarrow} (0) = a_0^2 .
\end{eqnarray}
Equation (\ref{a0equ}) has been made use of
in obtaining the preceding result.
The expression for $a_0$ is given in Eq. (\ref{a0}), with which
we obtain the final result of
Eq. (\ref{qian3D}).
Furthermore, it is straightforward to show, from
Eq. (\ref{gud(r)}), that
at small $r$,
\begin{eqnarray}  \label{gud(r)1}
g_{\uparrow \downarrow} (r)= g_{\uparrow \downarrow} (0)
+ \frac{\pi}{2} \lambda_3 g_{\uparrow \downarrow} (0) r .
\end{eqnarray}

\subsection{2D}

In 2D, one has,
\begin{eqnarray}   \label{gud2(r)}
g_{\uparrow \downarrow} (r)
= \biggl [ 1 - \lambda_2 \int_1^\infty  
d q \frac{1}{q} V_{eff} (q) J_0(qr) \biggr ]^2 .
\end{eqnarray}
Similar derivation to that in the 3D case leads to
\begin{eqnarray}
g_{\uparrow \downarrow} (0) = c_0^2 ,
\end{eqnarray}
or, by the use of Eq. (\ref{c0}), the final result of
Eq. (\ref{qian2D}). Furthermore, from Eq. (\ref{gud2(r)}),
one can obtain
\begin{eqnarray}  \label{gud(r)2}
g_{\uparrow \downarrow} (r)= g_{\uparrow \downarrow} (0)
+ 2 \lambda_2 g_{\uparrow \downarrow} (0) r ,
\end{eqnarray}

Evidently, the cusp condition of Eq. (\ref{cusp}) is satified
both in Eq. (\ref{gud(r)1}) and Eq. (\ref{gud(r)2}).
In fact, in Appendix B we shall show that, 
Eq. (\ref{cusp}) is satisfied, in general,
in the full ladder theory.

\begin{figure}
\unitlength1cm
\begin{picture}(5.0,6.0)
\put(-5.0,-4.0){\makebox(7.0,8.0){
\includegraphics{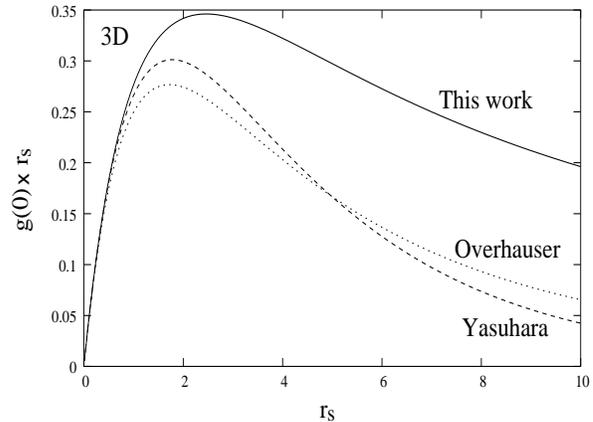}
}}
\end{picture}
\caption{On-top pair-correlation density multiplied by $r_s$ in
3D. The result of this work (Eq. (\ref{qian3D})) is compared with
Eq. (\ref{yasuhara}) (Yasuhara's formula 
in Refs. \cite{yasuhara2,yasuhara3,isawa}), and
Overhauser formula (Eq. (26) in Ref. \cite{overhauser}). 
}
\label{figure1}
\end{figure}

\section{Comparisons and discussions}

First of all, at limiting high density, we have, from
Eq. (\ref{qian3D}), 
\begin{eqnarray}  \label{highdensity}
g_{\uparrow \downarrow} (0) &=& 1- 2 \lambda_3  \nonumber  \\
&=& 1-0.663 r_s  ,
\end{eqnarray}
in 3D. Equation (\ref{highdensity}) is the 
same as the corresponding Yasuhara's result.\cite{yasuhara3}
We note that, the first order perturbation
calculation, \cite{burke,kimball,geldart,pople} which is believed
to approach to the exact result
at high density limit,
yields a result of $1- 0.7317 r_s$.
We plot $g(0) \times r_s$
calculated from Eq. (\ref{qian3D}) in Fig. 1, 
in comparison with that calculated
from Eq. (\ref{yasuhara}) \cite{yasuhara2,yasuhara3,isawa}. 
Notice
that the discrepancy between Eq. (\ref{qian3D}) 
and Eq. (\ref{yasuhara}), which appears not minor, arises purely
from the approximation of Eq. (\ref{kernel}) 
made in obtaining Eq. (\ref{yasuhara}) in Yasuhara's
theory.
In effect, Lowy and Brown \cite{brown} had thrown doubt on the validity
of the approximation of Eq. (\ref{kernel}). 
We hence justify their doubt, at least
for the limiting short range correlations. The result based on
Overhauser's proposal (Eq. (26) in Ref. \cite{overhauser})
is also shown in Fig. 1. The comparison hence indicates
that the coincidence between Overhauser's result (and the
corresponding numerical result of Gori-Giorgi and 
Perdew \cite{gori-giorgi1}) and Yasuhara's 
is accidental.

\begin{figure}
\unitlength1cm
\begin{picture}(5.0,6.0)
\put(-5.0,-4.0){\makebox(7.0,8.0){
\includegraphics{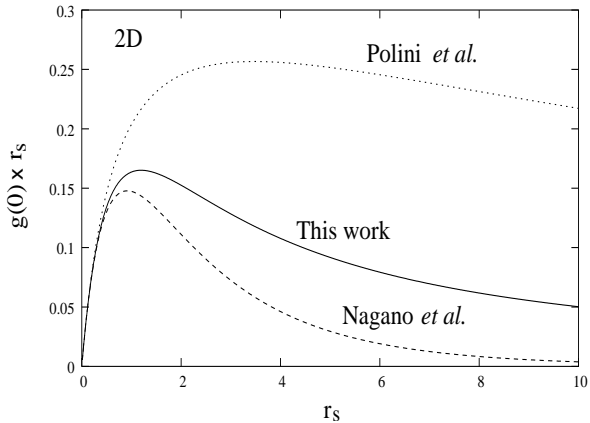}
}}
\end{picture}
\caption{On-top pair-correlation density multiplied by $r_s$ in 2D.
The result of this work (Eq. (\ref{qian2D}))
is compared with Eq. (\ref{nagano}) (Nagano {\em et al.} formula
in Ref. \cite{nagano}), and Polini {\em et al.} formula (Eq. (17)
in Ref. \cite{polini}).
}
\label{figure2}
\end{figure}
In Fig. 2, we plot $g(0) \times r_s$ in 2D
calculated from Eq. (\ref{qian2D}), together with
that from Eq. (\ref{nagano}).\cite{nagano} Once again, we emphasize
that the discrepancy is totally due to the approximation 
of Eq. (\ref{kernel}) made in obtaining Eq. (\ref{nagano}).
However, at limiting high density, both equations
yield the following same result:
\begin{eqnarray}
g_{\uparrow \downarrow} (0) = 1- 2 \lambda_2 .
\end{eqnarray}
For a comparison, we have also shown in Fig. 2 the result of
Eq. (17) in Ref. \cite{polini}, which was proposed
by Polini {\em et al.} based on an interpolation between
the first-order (second-order in terms of
the correlation energy) calculation for the weak-coupling limit
and Overhauser type calculation \cite{overhauser} for the
strong-coupling limit.

\section{Conclusions}

The proper approach to the short-range electron correlations 
in many-body theory is the ladder theory,
in which the effective potential between two scattering particles
satisfies
the Bethe-Goldstone equation of Eq. (\ref{BG1}).
In this paper, we have proved that, the ladder theory
satisfies the cusp condition for the pair-correlation density
in the homogeneous electron liquid.
This enhances our belief in the capability of
the ladder theory in describing the short-range correlations, especially
in calculating the pair-correlation density.

The main results obtained in this 
paper are, in effect, Eq. (\ref{qian3D})
and Eq. (\ref{qian2D}) given in the Introduction, 
in three dimensions and two dimensions
respectively, for the on-top
pair-correlation density in the homogeneous electron liquid.
These results have been derived by solving Eq. (\ref{BG2}),
in which the two scattering particles in the Bethe-Goldstone equation 
are approximately taken to be static. This approximation
should be reasonable since the
limiting short range structure of the pair-correlation is determined 
by the large transfer momentum behavior of the effective potential. 
The major theoretical progress made 
in this paper is that we have removed the
approximation of
Eq. (\ref{kernel}) frequently made 
in the literature for the Coulomb kernel in solving
Eq. (\ref{BG2}). Our solution to Eq. (\ref{BG2}) is thus exact.

\begin{acknowledgments}
This work was supported by the Chinese National Science Foundation
under Grant No. 10474001. 
\end{acknowledgments}

\appendix
 
\section{On the solutions to Eqs. (\ref{a0equ}) and (\ref{anequ}),
and Eqs. (\ref{c0equ}) and (\ref{cnequ}).}
                                                                                   
\begin{figure}
\unitlength1cm
\begin{picture}(5.0,6.0)
\put(-5.0,-4.0){\makebox(7.0,8.0){
\includegraphics{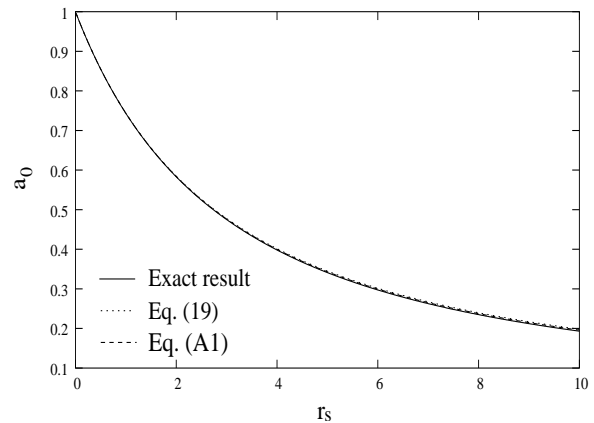}
}}
\end{picture}
\caption{$a_0$ calculated from Eq. (\ref{a0}) and
Eq. (\ref{a00}), together with the exact numerical result.
The inset shows the large $r_s$ region.
}
\end{figure}
\begin{figure}
\unitlength1cm
\begin{picture}(-1.4,0.0)
\put(-4.4,0.77){\makebox(7.0,8.0){
\includegraphics{a0inset.ps}
}}
\end{picture}
\label{figure3}
\end{figure}
A nearly exact solution for $a_0$ to
Eqs. (\ref{a0equ}) and (\ref{anequ}) has been given in Eq.
(\ref{a0}) in Sect. III by the truncation of
$a_n=0$ for $n \ge 3$. Below we give the solution for
$a_n$ by the truncation of $a_n =0$
for $n \ge 4$.
\begin{eqnarray}   \label{a00}
a_0 = \frac{175(14175 + 9585 \lambda_3
+ 2520 \lambda_3^2 + 256 \lambda_3^3)}{{\tilde D}_3} ,
\end{eqnarray}
\begin{eqnarray}  \label{a10}
a_1 = \frac{105 \lambda_3 (7875 + 2480 \lambda_3
+256 \lambda_3^2)}{{\tilde D}_3} ,
\end{eqnarray}
\begin{eqnarray}  \label{a20}
a_2= \frac{105 \lambda_3 (1575 + 2280 \lambda_3
+256 \lambda_3^2)}{{\tilde D}_3} ,
\end{eqnarray}
and
\begin{eqnarray}   \label{a30}
a_3= \frac{175 \lambda_3 (405 + 576 \lambda_3
+256 \lambda_3^2)}{{\tilde D}_3} ,
\end{eqnarray}
where
\begin{eqnarray}
{\tilde D}_3 = && 2480625 +4158000 \lambda_3
+2437200 \lambda_3^2  \nonumber \\
&& +634880 \lambda_3^3 +65536 \lambda_3^4 .
\end{eqnarray}
In Fig. 3, we plot the results for $a_0$ calculated
from Eqs. (\ref{a0}) and (\ref{a00}), together with
the corresponding exact numerical solution to Eqs (\ref{a0equ}),
and (\ref{anequ}). There is basically no difference
among them. We present the expressions of Eqs. (\ref{a00}),
(\ref{a10}), (\ref{a20}), (\ref{a30})
above for possible future reference.
                                                                                   
Similar expressions for the 2D case are given below.
\begin{eqnarray}  \label{c00}
c_0 = \frac{35(12288+6000 \lambda_2 + 1121 \lambda_2^2
+ 80 \lambda_2^3 )}{{\tilde D}_2} ,
\end{eqnarray}
\begin{eqnarray}
c_1 = \frac{1680 \lambda_2 (64 +14 \lambda_2
+ \lambda_2^2)}{{\tilde D}_2} ,
\end{eqnarray}
\begin{eqnarray}
c_2= \frac{105 \lambda_2 (192 +207 \lambda_2
+16 \lambda_2^2)}{{\tilde D}_2} ,
\end{eqnarray}
and
\begin{eqnarray}
c_3= \frac{350 \lambda_2 (24 +25 \lambda_2
+ 8 \lambda_2^2)}{{\tilde D}_2} ,
\end{eqnarray}
where
\begin{eqnarray}
{\tilde D}_2 =&& 430080 + 640080 \lambda_2 +290307 \lambda_2^2  \nonumber \\
&& + 55472 \lambda_2^3 +4096 \lambda_2^4 .
\end{eqnarray}
The corresponding illustration is given in Fig. 4.
\begin{figure}
\unitlength1cm
\begin{picture}(5.0,6.0)
\put(-5.0,-4.0){\makebox(7.0,8.0){
\includegraphics{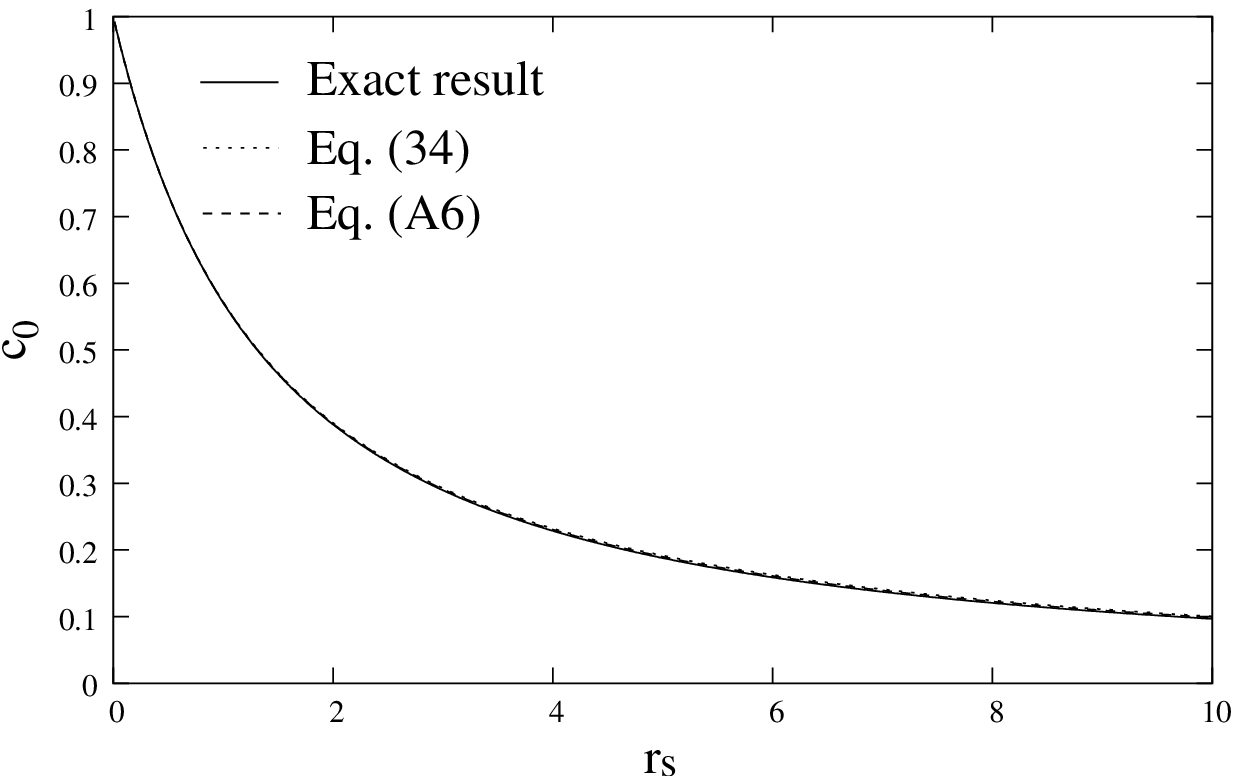}
}}
\end{picture}
\caption{$c_0$ calculated from Eq. (\ref{c0}) and
Eq. (\ref{c00}), together with exact numerical result.
The inset shows the large $r_s$ region.}
\end{figure}
\begin{figure}
\unitlength1cm
\begin{picture}(-1.4,0.0)
\put(-4.5,0.75){\makebox(7.0,8.0){
\includegraphics{c0inset.ps}
}}
\end{picture}
\label{figure4}
\end{figure}

\section{Cusp condition in the ladder theory}

Due to the singularity of the Coulomb potential
between electrons, the many-body schr\"odinger wavefunction
has a cusp when any two electrons coalesce.\cite{kato,bingel,pack}
This
fact leads to the cusp
condition of Eq. (\ref{cusp}) 
\cite{yasuhara1,ousaka,kimball,rajagopal,rassolov,sahni}
for the pair-correlation
density (also
known as Kimball relation
in the literature of many-electron theory).
Recently
it was claimed that Eq. (\ref{cusp}) is not
satisfied in the LT.\cite{CZ} In this appendix, we
give a rigorous proof for Eq. (\ref{cusp}) in the LT.
The proof will be formulated in 3D.

We start with the definition of the spin-parallel static
structure factor as follows:
\begin{eqnarray}
S_{\uparrow \downarrow} (q)= \frac{1}{N} \langle
{\hat n}_\uparrow (-{\bf q}) {\hat n}_\downarrow ({\bf q}) \rangle
-\frac{N}{2} \delta_{{\bf q}, 0},
\end{eqnarray}
where ${\hat n}_\sigma ({\bf q})$ is the spin-resolved density
operator and $N$ is the particle number.
It has been shown that the spin-antiparallel static structure factor
in the LT can be expressed in terms of the effective potential
of Eq. (\ref{BG1}) as \cite{yasuhara1,nagano}
\begin{eqnarray}  \label{Sud1}
S_{\uparrow \downarrow} (q) =&& \frac{1}{n}
{\sum_{{\bf p}, {\bf p}'}}'
\biggl [ 2 D({\bf p}, {\bf p}'; {\bf q})
V_{eff} ({\bf p}, {\bf p}'; {\bf q})   \nonumber \\
&& + \sum_{\bf k} D({\bf p}, {\bf p}'; {\bf k})
V_{eff}({\bf p}, {\bf p}'; {\bf k})   \nonumber \\
&& \times D({\bf p}, {\bf p}'; {\bf k} - {\bf q})
V_{eff}({\bf p}, {\bf p}'; {\bf k} - {\bf q}) \biggr ].
\end{eqnarray}

Next we examine the large momentum structure
of $S_{\uparrow \downarrow}(q)$.
For $p, p' \le k_F$ and $q \to \infty$, one has, from Eqs. (\ref{BG1})
and (\ref{D}), 
\begin{eqnarray}  \label{DV}
D({\bf p}, {\bf p}'; {\bf q}) && V_{eff} ({\bf p}, {\bf p}'; {\bf q})
= - \frac{1}{2 \epsilon_q} v(q)   \nonumber \\
\times \biggl [ 1 + && \sum_{\bf k} D({\bf p}, {\bf p}'; {\bf k})
V_{eff} ({\bf p}, {\bf p}'; {\bf k}) \biggr ],
\end{eqnarray}
which evidently goes to zero in the order of $O(1/q^4)$. Therefore
\begin{eqnarray}  \label{DVDV}
\sum_{\bf k} && D({\bf p}, {\bf p}'; {\bf k})
V_{eff} ({\bf p}, {\bf p}'; {\bf k})  \nonumber \\
&& \times D({\bf p}, {\bf p}'; {\bf q} - {\bf k})
V_{eff} ({\bf p}, {\bf p}'; {\bf q} - {\bf k})  \nonumber \\
&& = - \frac{1}{\epsilon_q} v(q)
\biggl [ 1 + \sum_{\bf k} D({\bf p}, {\bf p}'; {\bf k})
V_{eff} ({\bf p}, {\bf p}'; {\bf k}) \biggr ]  \nonumber \\
&& \times \sum_{{\bf k}'} D({\bf p}, {\bf p}'; {\bf k}')
V_{eff} ({\bf p}, {\bf p}'; {\bf k}').
\end{eqnarray}
In obtaining Eq. (\ref{DVDV}), we have used the following relation,
\begin{eqnarray}  \label{ff}
\lim_{q \to \infty} \sum_{\bf k} f({\bf k}) f({\bf q} - {\bf k})
= 2 f({\bf q})  \sum_{\bf k} f({\bf k}),
\end{eqnarray}
if $\lim_{q \to \infty} f({\bf q}) \sim  O(1/q^4)$. 
It seems that a mistake occurs in Ref. \cite{CZ}
due to a possible miss of the factor $2$ on the right hand side
of the preceding equation, as it was employed to derive Eq. (12) from
Eq. (11) in Ref. \cite{CZ}. 

Substituting Eqs. (\ref{DV}) and (\ref{DVDV}) into Eq. (\ref{Sud1}), one has
\begin{eqnarray}  \label{SUD}
S_{\uparrow \downarrow}(q) = - \frac{1}{n} \frac{v(q)}{\epsilon_q}
{\sum_{{\bf p}, {\bf p}'}}'
\biggl [ 1 + && \sum_{\bf k} D({\bf p}, {\bf p}'; {\bf k}) \nonumber \\
&& \times V_{eff} ({\bf p}, {\bf p}'; {\bf k}) \biggr ]^2.
\end{eqnarray}
On the other hand, from Eq. (\ref{gud1}), we have
\begin{eqnarray}   \label{gud0}
g_{\uparrow \downarrow}(0)
= \frac{4}{n^2} {\sum_{{\bf p}, {\bf p}'}}' \biggl [ 1 +
\sum_{\bf q} && D({\bf p}, {\bf p}'; {\bf q})     \nonumber \\
&& \times V_{eff}({\bf p}, {\bf p}'; {\bf q}) \biggr ]^2.
\end{eqnarray}
Comparing Eq. (\ref{SUD}) and Eq. (\ref{gud0}) yields
\begin{eqnarray}
S_{\uparrow \downarrow}(q) = - \frac{2 \pi e^2 n m}{q^4} 
g_{\uparrow \downarrow}(0) .
\end{eqnarray}
Combining the preceding result with the following well-known
relation \cite{kimball},
\begin{eqnarray}
\lim_{q \to \infty} q^4 S_{\uparrow \downarrow}(q) = - 2 \pi n \frac{
\partial g_{\uparrow \downarrow}(r)}{\partial r}|_{r=0},
\end{eqnarray}
one proves the cusp condition of Eq. (\ref{cusp}) in the LT.

The above proof can be straightforwardly extended to the 2D case. In fact, in
2D, it can be similarly shown 
\begin{eqnarray}
S_{\uparrow \downarrow}(q) = - \frac{\pi e^2 n m}{q^3}
g_{\uparrow \downarrow}(0),
\end{eqnarray}
in the LT.
Combining the above result with the following relation \cite{qian}
\begin{eqnarray}
\lim_{q \to \infty} q^3 S_{\uparrow \downarrow}(q) = - \frac{1}{2} 
\pi n \frac{
\partial g_{\uparrow \downarrow}(r)}{\partial r} |_{r=0},
\end{eqnarray}
leads to Eq. (\ref{cusp}) for the 2D case.

\end{document}